\documentclass[epj]{svjour}
\usepackage{graphics}
\begin{document}
\title{\boldmath{$K^{0}\pi^{0}\Sigma^{+}$ and $K^{*0}\Sigma^{+} $} photoproduction off the proton}
\author{
  M.~Nanova\inst{1},
  J.~C.~S.~Bacelar\inst{2},
  B.~Bantes\inst{3},
  O.~Bartholomy\inst{4},
  D.~Bayadilov\inst{4,5},
  R.~Beck\inst{4},
  Y.A.~Beloglazov\inst{5},
  R.~Castelijns\inst{2,a},
  V.~Crede\inst{6},
  H.~Dutz\inst{3},
  A.~Ehmanns\inst{4},
  D.~Elsner\inst{3},
  K.~Essig\inst{4},
  R.~Ewald\inst{3},
  I.~Fabry\inst{4},
  K.~Fornet-Ponse\inst{3},
  M.~Fuchs\inst{4},
  Ch.~Funke\inst{4},
  R.~Gothe\inst{3,b},
  R.~Gregor\inst{1},
  A.~B.~Gridnev\inst{5},
  E.~Gutz\inst{4},
  P.~Hoffmeister\inst{4},
  I.~Horn\inst{4},
  I.~Jaegle\inst{7},
  J.~Junkersfeld\inst{4},
  H.~Kalinowsky\inst{4},
  S.~Kammer\inst{3},
  V.~Kleber\inst{3},
  Frank~Klein\inst{3},
  Friedrich~Klein\inst{3},
  E.~Klempt\inst{4},
  M.~Konrad\inst{3},
  M.~Kotulla\inst{1},
  B.~Krusche\inst{7},
  M.~Lang\inst{4},
  J.~Langheinrich\inst{3,b},
  H.~L\"ohner\inst{2},
  I.V.~Lopatin\inst{5},
  J.~Lotz\inst{4},
  S.~Lugert\inst{1},
  D.~Menze\inst{3},
  J.~G.~Messchendorp\inst{2}, 
  T.~Mertens\inst{7},
  V.~Metag\inst{1},
  C.~Morales\inst{3},
  D.V.~Novinski\inst{5}, 
  R.~Novotny\inst{1}, 
  M.~Ostrick\inst{3,c},
  L.~M.~Pant\inst{1,d}, 
  H.~van Pee\inst{4}, 
  M.~Pfeiffer\inst{1},
  A.~Radkov\inst{5}, 
  A.~Roy\inst{1,e}, 
  S.~Schadmand\inst{1,a},
  Ch.~Schmidt\inst{4}, 
  H.~Schmieden\inst{3}, 
  B.~Schoch\inst{3}, 
  S.~V.~Shende\inst{2},
  V.~Sokhoyan\inst{4}, 
  A.~S\"ule\inst{3}, 
  V.~V.~Sumachev\inst{5}, 
  T.~Szczepanek\inst{4},
  U.~Thoma\inst{1,4}, 
  D.~Trnka\inst{1}, 
  R.~Varma\inst{1,e},
  D.~Walther\inst{3,4}, 
  Ch.~Weinheimer\inst{4,f}, 
  \and Ch.~Wendel\inst{4}\\ 
(The CBELSA/TAPS Collaboration)
\mail{{Mariana.Nanova@physik.uni-giessen.de}}
}
\titlerunning{$\Sigma^{+}$(1189) photoproduction off the proton}
\authorrunning{M.~Nanova et al.}
\institute{%
  $^1$II. Physikalisches Institut, Universit\"at Gie{\ss}en, Germany\\
  $^2$Kernfysisch Versneller Institut, Groningen, The Netherlands\\
  $^3$Physikalisches Institut, Universit\"at Bonn, Germany\\
  $^4$\mbox{Helmholtz-Institut f\"ur Strahlen- u. Kernphysik, Universit\"at Bonn, Germany}\\
  $^5$Petersburg Nuclear Physics Institute, Gatchina, Russia\\
  $^6$Department of Physics, Florida State University, Tallahassee, FL, USA\\
  $^7$Physikalisches Institut, Universit\"at Basel, Switzerland\\
  $^a$Present address: Forschunszentrum J\"ulich, Germany\\
  $^b$Present address: University of South Carolina, Columbia, SC, USA\\
  $^c$Present address: Universit\"at Mainz, Germany\\
  $^d$Present address: BARC, Mumbai, India \\
  $^e$Present address: I.I.T. Powai, Mumbai, India\\
  $^f$Present address: Universit\"at M\"unster, Germany\\
}
\date{Received: date / Revised version: date}
\abstract{The exclusive reactions $\gamma p \rightarrow
  K^{*0} \Sigma^+(1189)$ and $\gamma p \rightarrow  K^{0}
  \pi^{0}\Sigma^+(1189)$, leading to the p 4$\pi^{0}$ final state, have been 
  measured with a tagged photon beam for incident energies from threshold up to
  2.5 GeV. The experiment has been performed at the tagged photon facility of
  the ELSA accelerator (Bonn). The Crystal Barrel and TAPS detectors were
  combined to a photon detector system of almost 4$\pi$ geometrical
  acceptance. Differential and total cross sections are reported. At
  energies close to the threshold, a flat angular distribution has
  been observed for the reaction $\gamma p\rightarrow  K^{0} \pi^{0}\Sigma^+$
  suggesting dominant $s$-channel production.  $\Sigma^*(1385)$ and
  higher lying hyperon states have been observed. An enhancement in the
  forward direction in the angular distributions of the reaction $\gamma p
  \rightarrow K^{*0}\Sigma^+$ indicates a $t$-channel exchange contribution
  to the reaction mechanism. The experimental data are in reasonable agreement
  with  recent theoretical predictions.
}
\maketitle
\section{Introduction}
\label{intro}
\begin{sloppypar}
The internal structure of the nucleon is reflected in the rich pattern of
baryon resonances. The number of experimentally observed resonances is
much smaller than predicted from theory~\cite{Capstick}. This is often
referred to as the 'missing' resonance problem. Baryon resonances
have often large widths and overlap largely, which makes the study of the
excited states particularly difficult. It is possible to overcome this problem
by looking at specific decay channels. Up to now most existing data are based
on elastic $\pi N$ scattering experiments. If the hypothesis is correct that
the missing states are unobserved because they couple weakly to the $\pi N$
decay channel, it may be possible to establish some of these missing states in
other channels. Some of the resonances are predicted to decay into final
states with strange particle pairs, coupling strongly to $K\Lambda$ and
$K\Sigma$~\cite{Capstick1}. Strangeness production experiments will therefore
be an important tool to establish 'missing' resonances and their
properties or to disprove their existence. Recently new measurements of
differential and total cross sections of hyperon photoproduction have been
reported by the SAPHIR~\cite{lawall}, LEPS~\cite{sumi},
CBELSA/TAPS~\cite{ralph} and CLAS~\cite{brad,brad1} collaborations. A partial
wave analysis  provided a good description of these data by introduction of a
new state with $P_{13}$ quantum numbers and two solutions for the mass -
either at 1885 MeV or at 1970 MeV ~\cite{anis1,anis2}.\\ 
Higher mass nucleon resonances could favour decays into
$K^{*}\Sigma$. Therefore $K^{*}$ vector meson photoproduction can be used to
search for nucleon resonances which couple strongly to the $K^{*}Y$
channel, where $Y$ denotes a hyperon ~\cite{Capstick2}.
On the other hand, photoproduction of $K^{*}$ shares elements in common with
other strangeness production reactions, such as $\gamma p \rightarrow
K\Lambda$ and $\gamma p \rightarrow K \Sigma$ or $\gamma p \rightarrow \pi K
\Sigma$, which lead to $N^{*}$ or to $N^{*}$ and $\Delta^{*}$ resonance
excitations with different couplings. The investigation of higher
lying hyperon $\Sigma^{*}$ resonances will provide more information about the
baryon resonances in meson-hyperon decay channels and help to understand their
contribution as background to $K^*$ photoproduction
in the reaction  $\gamma p \rightarrow K^{*0} \Sigma^+$. \\
The cross sections for these reaction channels are small and their experimental
identification is difficult. Therefore their investigation became
only feasible when high quality photon beam facilities combined with 4$\pi$
high resolution detectors became available. Here we present the experimental
data of $\Sigma^+$(1189) photoproduction off the proton by analysing:\\
(1) $\gamma p \rightarrow \pi^{0} K^{0}\Sigma^{+} \rightarrow \pi^{0}
(\pi^{0} \pi^{0})(\pi^{0} p) \rightarrow 8\gamma p $\\
This reaction contains the following isobars:\\
(2) $\gamma p \rightarrow K^{*0}\Sigma^{+}$\\
(3) $\gamma p \rightarrow K^{0}\Sigma^{*+}$\\
(4) $\gamma p \rightarrow (\pi^{0} K^{0}\Sigma^{+})_{n.r.}$\\
where the $K^{*0}$(892) decays into $K^{0}\pi^{0}$ and the $\Sigma^{*+}$ decays
into $\Sigma^{+}\pi^{0}$. The contribution of a non-resonant (n.r.) $
K^{0}\pi^{0}$ pair in (4) occurs together with a $\Sigma^{+}$(1189).\\
Recent theoretical studies of $\Sigma^+$ photoproduction in the channel
$\gamma p \rightarrow \pi^{0} K^{0} \Sigma^{+}$ ~\cite{oset} have been
performed using a chiral unitary approach for meson-baryon scattering in the
energy range close to 1700 MeV, below the threshold for $K^{*}$ production. 
The theoretical model is
based on the assumption that the $\Delta^{*}$(1700) is excited and decays into
$K\Sigma^{*}$(1385) or $\eta\Delta$(1232). It has been applied to calculate the
cross sections of the reactions $\gamma p \rightarrow p \pi^{0} \eta$ and
$\gamma p \rightarrow \pi^{0} K^{0}\Sigma^+$  ~\cite{oset1}. The main
conclusion is that the mechanism of both reactions is similar - going through
the production of the $\Delta^{*}(1700)$, which is dynamically generated with
strong couplings to the $\eta\Delta$ and $K\Sigma^{*}$. The current data will
be compared with the predictions of this model for the reaction $\gamma p
\rightarrow \pi^{0} K^{0}\Sigma^+$.\\ 
Quark model predictions for $K^{*}$
photoproduction via nucleon resonance excitations in the channels $\gamma
p \rightarrow K^{*0} \Sigma^+$ and $\gamma p \rightarrow  K^{*+}
\Sigma^{0}$ were presented in Ref.~\cite{zhao}. In the model,
resonances are treated as genuine quark states. There are only two free
parameters corresponding to the vector and tensor couplings which depend on
the quark mass. Using this approach the cross sections for $K^*$ production
have been predicted based on SU(3) symmetry and quark coupling parameters
extracted from non-strange production like $\omega$ meson production. The
assumption of {\it t}-channel $K$ exchange in this model leads to strong
forward peaking of $K^{*0}$ at higher energies ($>$ 2 GeV).\\
Another theoretical prediction for a $t$-channel exchange dominated reaction
mechanism in $K^{*}$ photoproduction involves the assumption that the
scalar $\kappa$(800) meson may play an important role ~\cite{Oh}. It was
demonstrated that $K$ exchange could describe the reaction mechanism in
$K^{*}\Lambda$ production, but for the $K^{*}\Sigma$ production the
contribution from the $\kappa$(800) meson could
be substantial.
The results of the CLAS collaboration have also been compared to
this theoretical prediction and are in good agreement. Nevertheless, the open
question here is the controversial structure of the $\kappa$(800)
meson. We will compare our data to this model too.\\
Experimentally, $K^{*+}(892)$ photoproduction has been studied with the SAPHIR
detector, in the reaction $\gamma p \rightarrow K^{*+}
\Lambda$~\cite{wieland}. The $K^{*+}$ was reconstructed from $K^{0} \pi^{+}$
where the measured charged particles are $\pi^{\pm}$ and proton. The
differential cross sections show a forward peaking of the $K^{*+}$ meson. The
measured total cross section is 0.35 $\mu$b at 2.2 GeV incident photon
energy.\\ In recent studies of $K^{*0}$ photoproduction  by the CLAS
collaboration in the reaction $\gamma p \rightarrow K^{*0}
\Sigma^+$~\cite{Hleiq}, the $K^{*0}$ was reconstructed from the
detected particles $K^{+}$ and $\pi^{-}$; the $\Sigma^+$ was treated as
missing particle. The angular distributions are forward peaked; good
agreement with the quark model of ~\cite{zhao} was achieved after a slight
adjustment of the vector and tensor $K^{*}$ couplings to the nucleon. The
small enhancement of the cross section at backward
angles has been interpreted as effect of $s$- and $u$-channel resonances that
couple to $K^{*0}\Sigma^{+}$. The production of higher hyperon resonances
$Y^{*}$, decaying into $\Lambda \pi$ or $\Sigma \pi$, has overlapping
kinematics with $K^{*}$ production leading to a background for the channel
$\gamma p \rightarrow K^{*0} \Sigma^+$.\\
Our data presented here will be compared with predictions of the available
theoretical models and with the published experimental results. The reactions
(1)-(4) have been identified via the neutral decay channels, which exclude
the contamination from hyperon resonances subsequently decaying via
$\Lambda$ production. The contribution from higher $\Sigma^{*}$ hyperon
resonances, decaying into $\Sigma\pi$, provide an important background to the
reaction (2). We will show evidence for identified higher hyperon states. For
photon energies above 1850 MeV, $\Sigma$ production is dominated by
reaction (2). Thus it is important to identify the $\Sigma^*$ contribution
against the leading $K^*$ contribution, since they are both leading to the
same final state.\\ 
This paper is organized as follows: In Section 2 we describe the
experiment. Section 3 provides the analysis method and event
reconstruction. Section 4 shows how to reconstruct and
remove the p$\pi^0 \eta$ events which have the same final state and
represent a considerable background to the reactions of interest. In Section 5
we discuss the reconstruction of $K^{0}$, $K^{*0}$ and $\Sigma^+$. The
differential and total cross sections are given and discussed in Section 6.
The paper is summarised in Section 7. 
\end{sloppypar}
\begin{figure*} 
   \resizebox{1.\textwidth}{!}{%
       \includegraphics{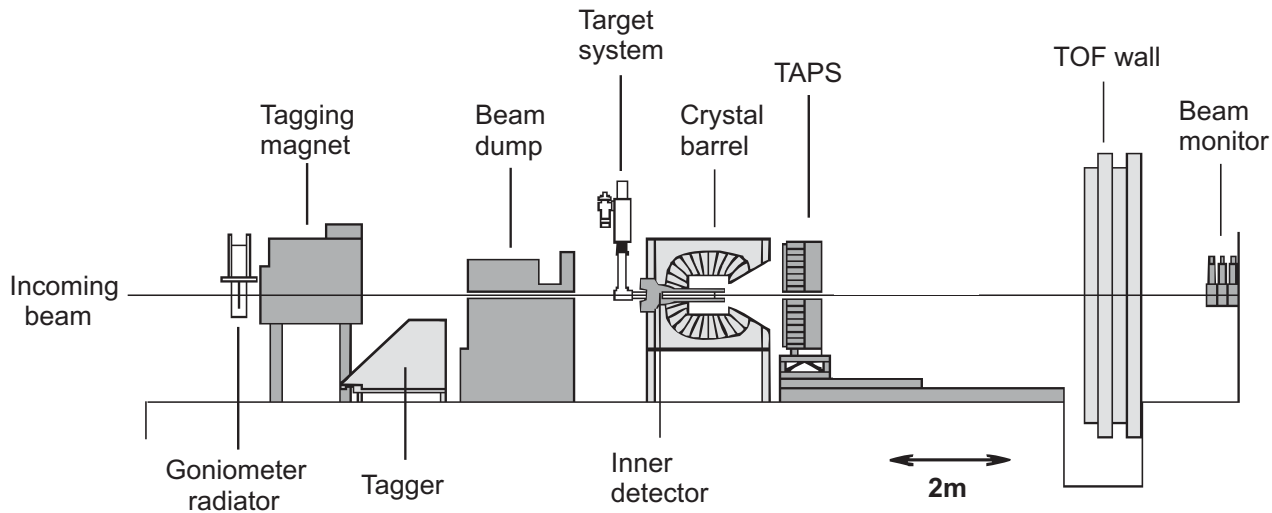}}
  \caption{Sketch of the experimental setup CBELSA/TAPS. The electron beam
  enters from the left side, hits the radiator and produces
  bremsstrahlung. The photons are energy tagged and hit the LH$_2$ target,
  which is in the center of the Crystal Barrel detector. The TAPS detector is
  placed just after the CB and serves as a forward wall of the CB. Charged
  particles leaving the target are identified in the inner scintillating-fibre
  detector  and in the plastic scintillators in front of each BaF$_2$ crystal
  in TAPS. A photon counter for the flux determination ($\gamma$ intensity
  monitor) is placed further downstream. \label{exp}} 
\end{figure*}
\section{The experiment}
\label{sec:1}
Data have been taken with the detector systems Crystal Barrel
(CB)~\cite{aker92} and TAPS~\cite{Novotny1,Gabler1} at the 3.5 GeV electron
stretcher facility ELSA~\cite{Husmann,Hillert}. The detector setup is shown
schematically in figure~\ref{exp}. Electrons extracted from ELSA with energy
$E_0$ hit a primary radiation target, a thin copper or diamond crystal, and
produce bremsstrahlung ~\cite{Elsner}. The tagging system consists of 480
scintillating fibers and 14 partly overlapping scintillator bars. It provides
the corresponding energy of the photons ($E_\gamma = E_0 - E_e^{-}$)
from the deflection of the scattered electrons in a magnetic field. Photons
were tagged in the energy range from 0.5 GeV up to 2.9 GeV
for an incoming electron energy of 3.2 GeV. The total tagged photon intensity
was about $10^{7}$ $s^{-1}$ in this energy range. The energy resolution varied
between 2 MeV for the high photon energies and 25 MeV for the low photon
energies at an electron beam energy of 3.2 GeV. The part of the beam that did
not produce any bremsstrahlung photons is deflected by the magnet as
well. Since the electrons have retained their full energy the curvature of
their track is smaller and they pass over the tagger into a beam dump.\\
At the end of the beam line a $\breve{C}$erenkov detector consisting of 9 lead
glass crystals has been installed (fig.~\ref{exp}) which measures those
photons that pass through the target without undergoing an interaction. The
information provided from this detector has been used for the photon flux
determination (see section 6.1). \\
The Crystal Barrel detector, a photon calorimeter consisting of 1290 CsI(Tl)
crystals ($\approx$16 radiation lengths), covered the complete azimuthal angle
and the polar angle from $30^o$ to $168^o$. The liquid hydrogen target in the 
center of the CB (5 cm in length, 3 cm in diameter) has been surrounded by a
scintillating fibre-detector to detect charged particles ~\cite{suft}. The CB
has been combined with a forward detector - the TAPS calorimeter - consisting
of 528 hexagonal BaF$_2$ crystals ($\approx$12 $X_0$), covering polar angles
between $5^o$ and $30^o$ and the complete 2$\pi$ azimuthal angle. In front of
each BaF$_2$ module a 5 mm thick plastic scintillator has been mounted for the
identification of charged particles. The combined CB/TAPS detector covered
99\% of the full 4$\pi$ solid angle. The high granularity of this system makes
it very well suited for the detection of multi-photon final states.\\
The first level trigger was derived from TAPS, requiring either one or two
hits above different thresholds. The second level trigger was based on a fast
cluster recognition (FACE) logic, providing the number of clusters in the
Crystal Barrel. For part of the data the minimal number of hits in FACE
was one, otherwise at least two hits were requested, which did not
introduce any bias for the channels analyzed here.
\begin{figure} 
\resizebox{0.5\textwidth}{!}{%
  \includegraphics{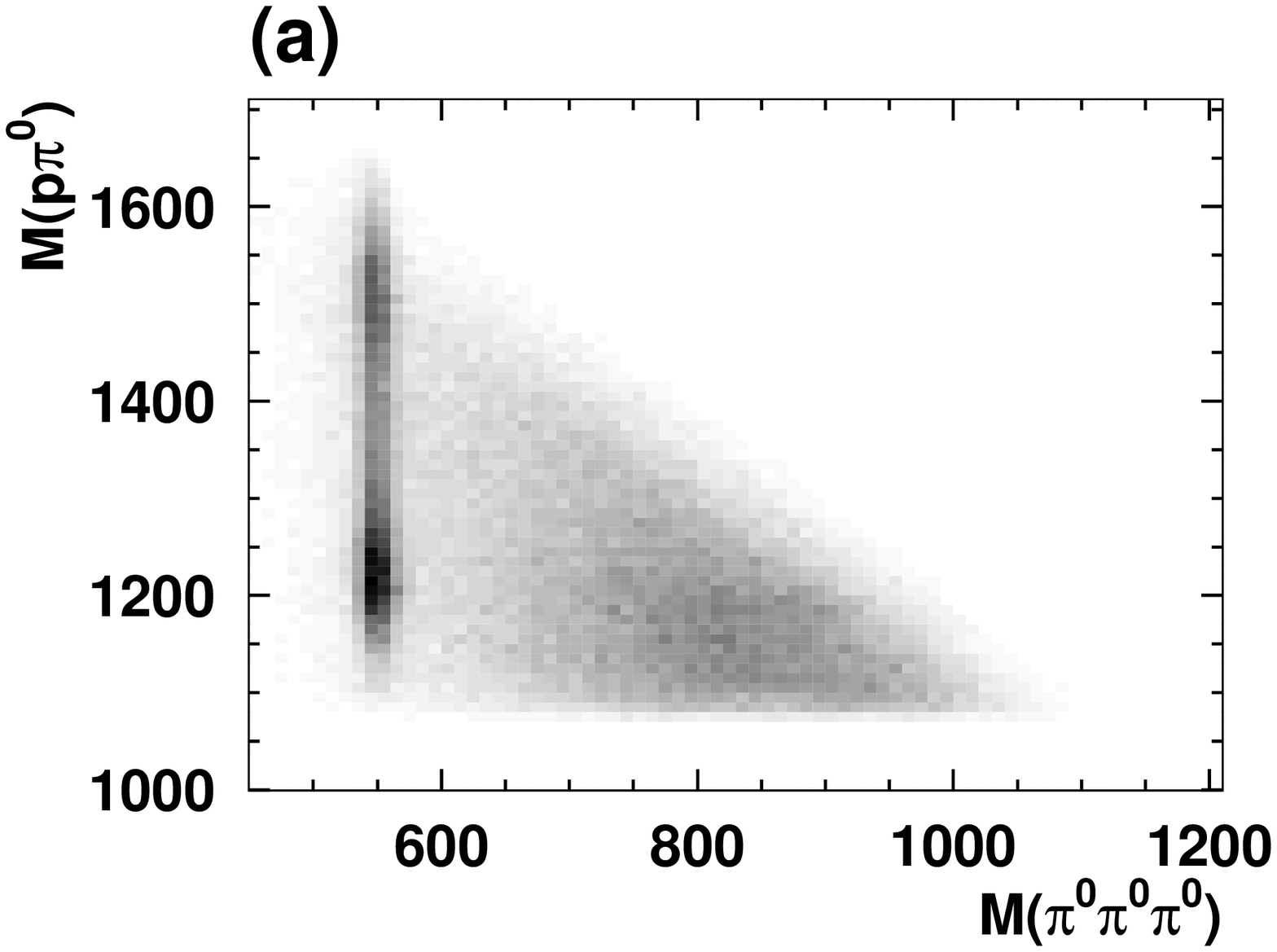}}
\resizebox{0.5\textwidth}{!}{%
  \includegraphics{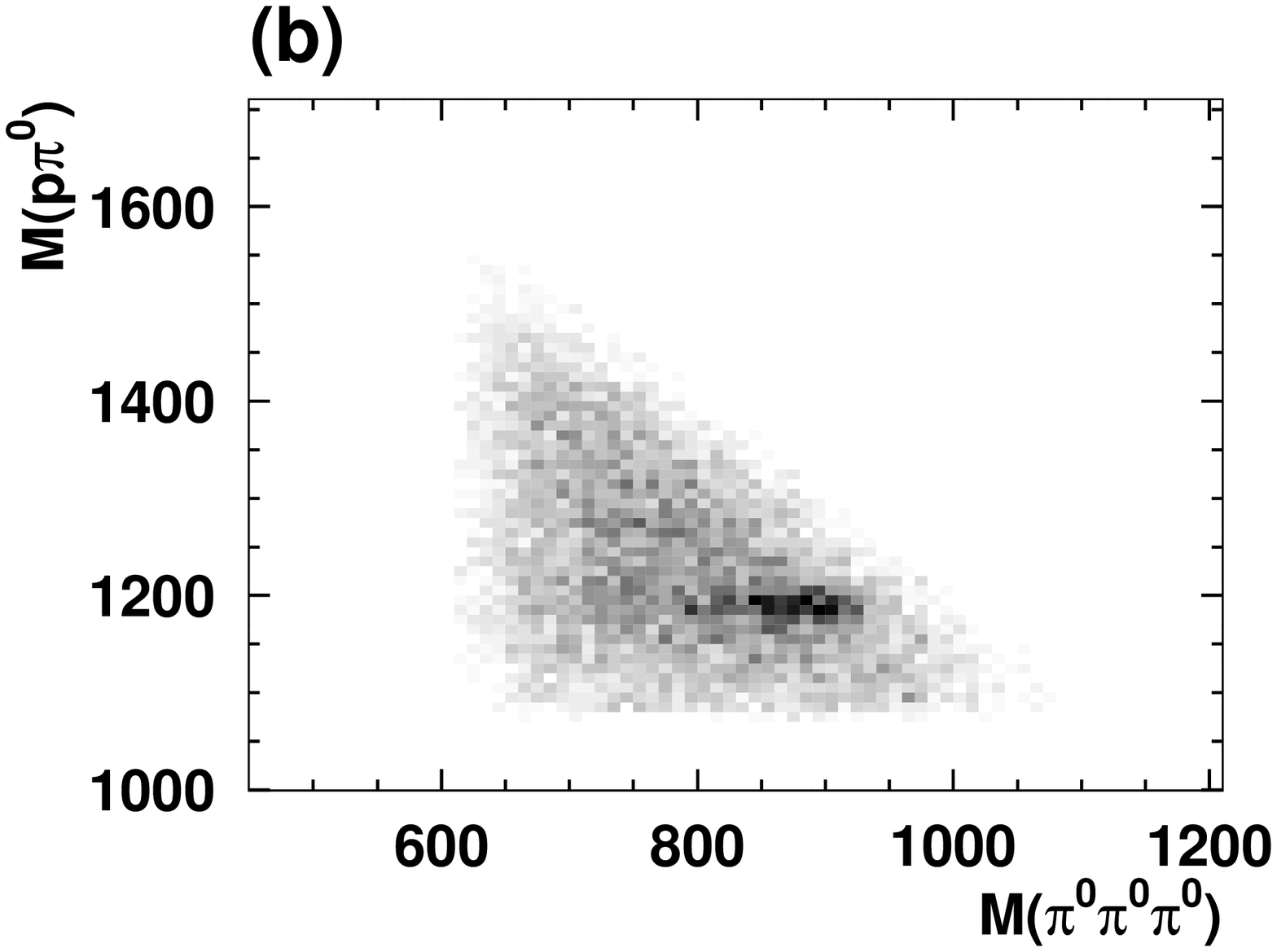}}
\caption{(a): Invariant mass M(p$\pi^{0}$) against M($\pi^{0}\pi^{0}\pi^{0}$)
  in the energy range 2000-2150 MeV, after applying the kinematic fit. The
  $\eta$ peak is clearly seen at 547 MeV. The resolution ($\sigma$) in
  M($\pi^{0}\pi^{0}\pi^{0}$) is 9 MeV. (b): After eliminating the events with
  M($\pi^{0}\pi^{0}\pi^{0}$) $<$ 600 MeV, a peak at M($\pi^{0}\pi^{0}\pi^{0}$)
  = 896 MeV and M(p$\pi^{0}$) = 1189 MeV is observed, showing correlated
  $K^*\Sigma$ production. \label{eta}}
\end{figure}
\section{Event reconstruction and event selection}
\label{sec:2}
The events due to the reactions $\gamma p \rightarrow K^{*0}\Sigma^{+}$ and
$\gamma p \rightarrow K^0\pi^0\Sigma^{+}$ were reconstructed from the measured
eight photons and the proton in the final channel. Only events containig
exactly nine clusters - eight neutral and one additional charged hit - were
selected. The charged clusters were identified by using the plastic
scintillators in front of the TAPS detector and the fibre detector in the
CB. In order to reduce the background, a cut in a missing mass spectrum
derived from the identified eight photons was applied. The cut selected events
in the region of the nucleon mass; the width of the missing mass cut varied as
a function of the incident photon energy (40
MeV at $E_\gamma$ = 1 GeV to 120 MeV at $E_\gamma$ = 2.6 GeV). Events
which survived this cut were kinematically fitted to the hypothesis
$\gamma p \rightarrow p_{miss} \pi^{0}\pi^{0}\pi^{0}\pi^{0}$. The procedure is
described in detail in~\cite{Pee}. 
Measurements of the deposited energy and the direction of the photons in the CB
and TAPS calorimeters were used in the fit. For the proton, only the two
angles of its trajectory were used, its energy was calculated, since
the fit is overconstrained. The constraints applied in
this analysis are energy and momentum conservation and the invariant masses of
the pions.\\
\begin{figure} 
\resizebox{0.6\textwidth}{!}{%
  \includegraphics{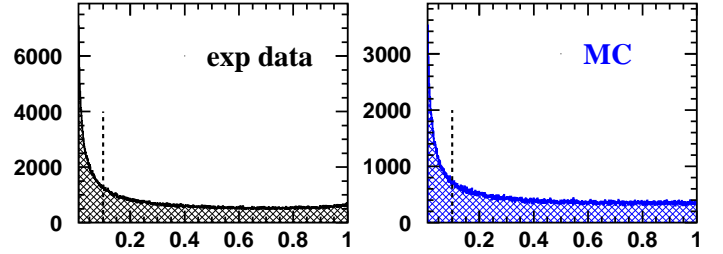}}
\caption{Confidence level distributions for the kinematic fit to the
  hypothesis $\gamma p \rightarrow p_{miss} \pi^{0}\pi^{0}\pi^{0}\pi^{0}$ for
  experimental data (left) and the corresponding Monte Carlo simulation
  (right). The dashed verticle line shows the applied cut on the confidence
  level. \label{conf}}
\end{figure}
The confidence level distribution of the fit is shown in
fig.~\ref{conf}. Above 20$\%$ the distribution is flat in both data and Monte
Carlo events. The combination of $\gamma$ pairs yielding the highest
confidence level was taken as the correct one. Events for which the kinematic
fit yielded a confidence level of less than 10$\%$ were removed from the data.
In order to eliminate time accidental background a prompt coincidence
between a photon in TAPS and an electron in the tagger was required. Random
time coincidences were subtracted, using events outside
the prompt time coincidence window. More details of this procedure can be
found in ~\cite{david}.
\section{The reaction $\gamma p \rightarrow p \pi^0 \eta$ }
\label{sec:3}
An important competing channel, leading to the same final state is
$\gamma p\rightarrow p \pi^{0}\eta\rightarrow
p4\pi^{0}$. Calculations~\cite{oset1} predicted 30 times larger  cross section
in comparison to the reactions with $\Sigma^{+} $ production. Experimentally,
the cross section for $\gamma p\rightarrow p \pi^{0}\eta$ reaction was
determined to $\sim$4$\mu$b~\cite{horn}. Events with one $\pi^0$ and 
three $\pi^{0}$'s from $\eta$-decay were selected. The two-dimensional plot of
the $p\pi^0$ invariant mass versus that of the 3$\pi^{0}$ system as a
result of the 4 possible combinations is shown in fig.~\ref{eta}a. A vertical
band for $\eta$ around 547 MeV can be seen. The $p\pi^0$ invariant mass
distribution within the $\eta$ band exhibits a strong peak due to the
$\eta\Delta(1232)$ intermediate state. The results from the analysis of
the $p\pi^0\eta$ channel will be published separately. In the present analysis
events due to $p\pi^0\eta$ were removed by a cut M($\pi^{0}\pi^{0}\pi^{0}$)
$<$ 600 MeV (fig.~\ref{eta}b). After these cuts, 9500 $p4\pi^{0}$ events remain
for the analysis of strangeness production.      
\begin{figure*} 
\resizebox{0.35\textwidth}{!}{%
  \includegraphics{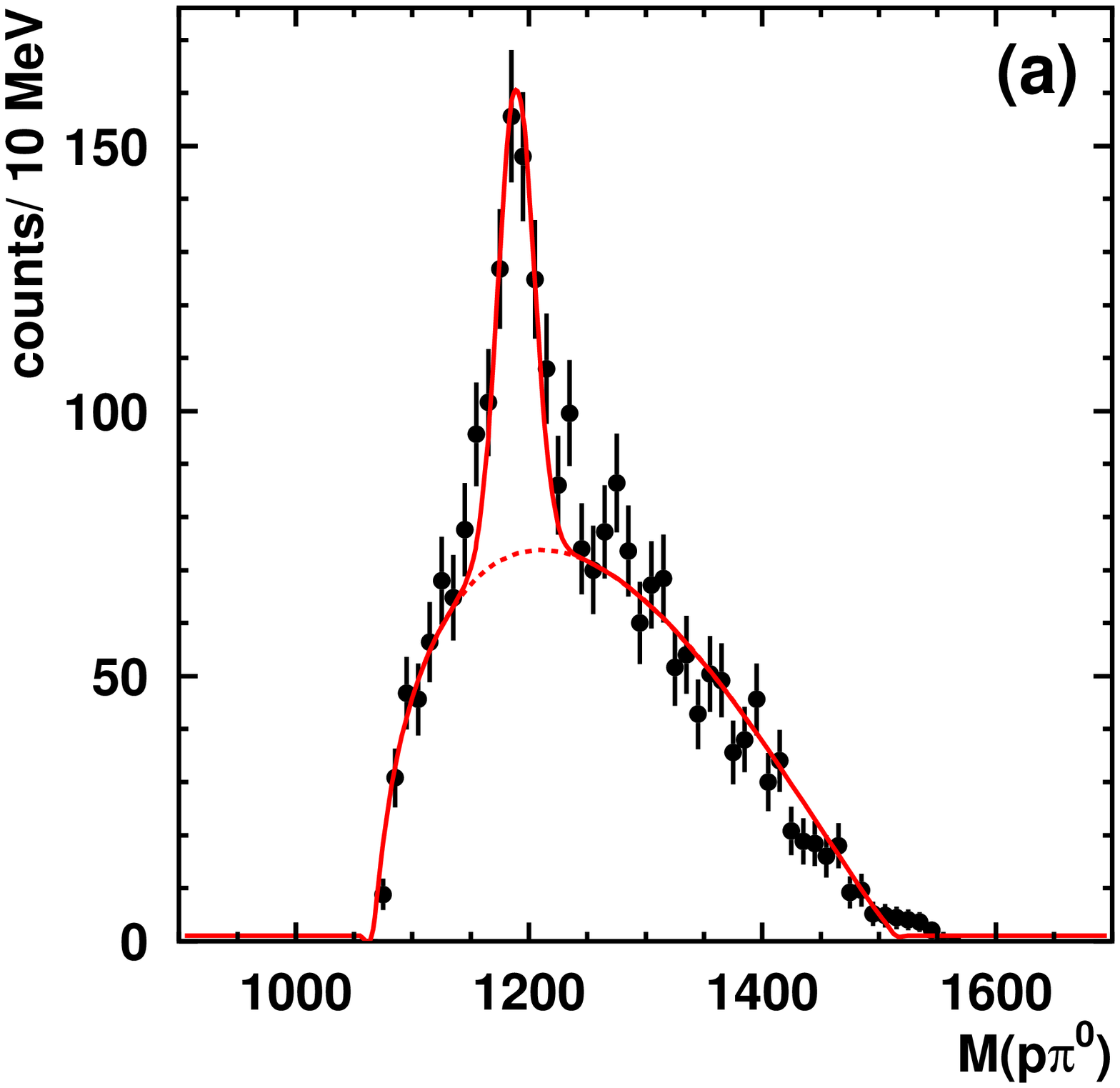}}
\resizebox{0.35\textwidth}{!}{%
  \includegraphics{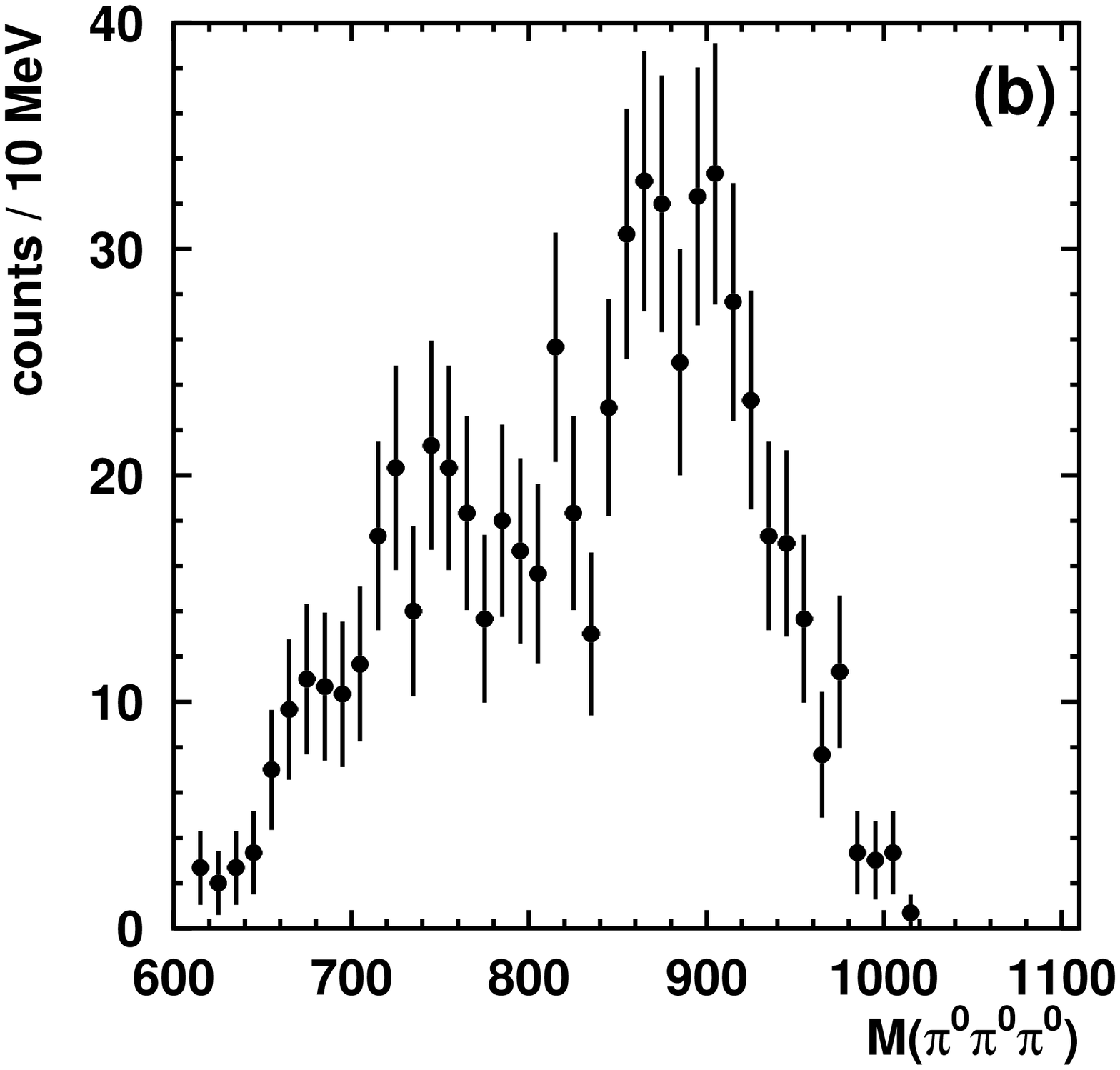}}
\resizebox{0.35\textwidth}{!}{%
  \includegraphics{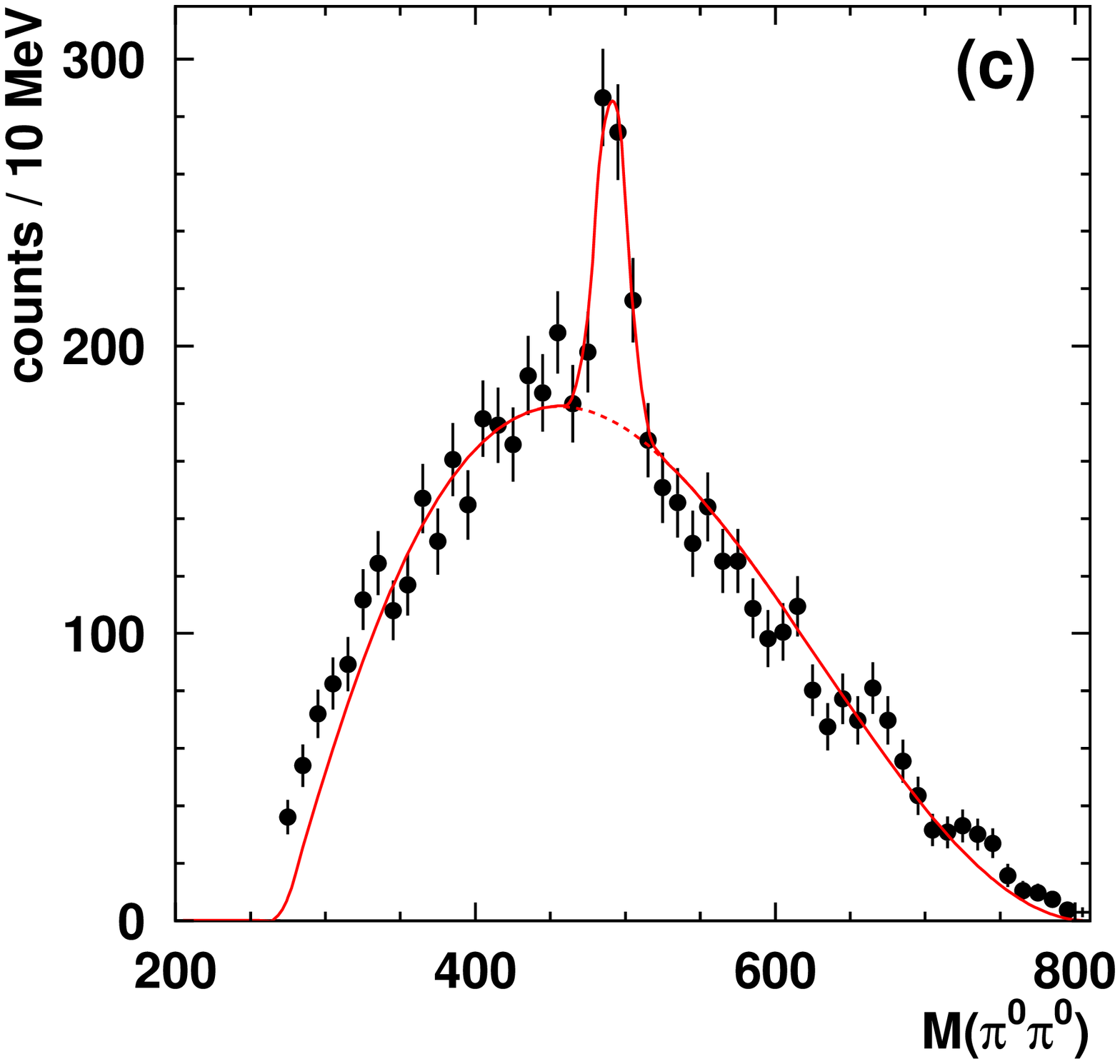}}
\caption{Projections of fig.~\ref{eta}b on the y-axis (a) and (b) on the x-axis
  with a cut of 1160-1220 MeV on the p$\pi^{0}$ invariant mass spectrum
  (see fig.~\ref{eta}). The incoming photon energies are between 2000 and
  2150 MeV. The solid curve shows the fit of the signal and the
  background. There are 340 events assigned to $\Sigma^+$ and 100 events due
  to $K^{*0}$. A resolution of 15 MeV ($\sigma$ ) for $\Sigma^{+}$ is given by
  the fit. The $\pi^{0}\pi^{0}$ invariant mass
  spectrum from all combinations of 4$\pi^{0}$-events, cutting on the
  $p\pi^{0}$ invariant mass spectrum from 1160-1220 MeV, is shown in (c). The
  solid curve shows the fit to the background and the $K^{0}$ signal. The fit
  yields a $K^{0}$ mass resolution of 10 MeV ($\sigma$). $\Sigma^{+}$ (a) and
  $K^{0}$ (c) yield agree within 10\% . \label{sigma}}
\end{figure*}
\section{The yield of hyperon $\Sigma^{+}$, $K^{0}$ and $K^{*0}$ mesons }
\label{sec:5}
\begin{figure} [h]
\resizebox{0.5\textwidth}{!}{%
  \includegraphics{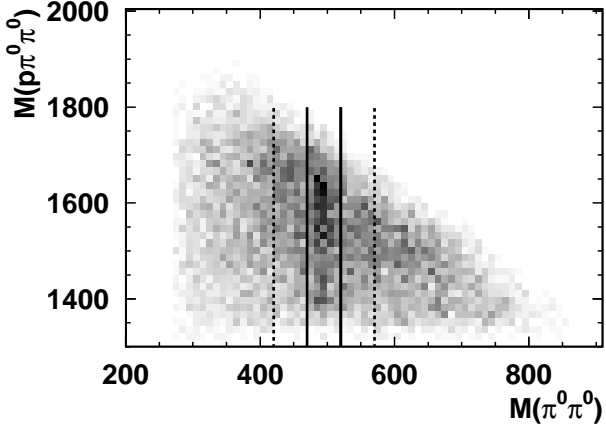}}
\resizebox{0.5\textwidth}{!}{%
  \includegraphics{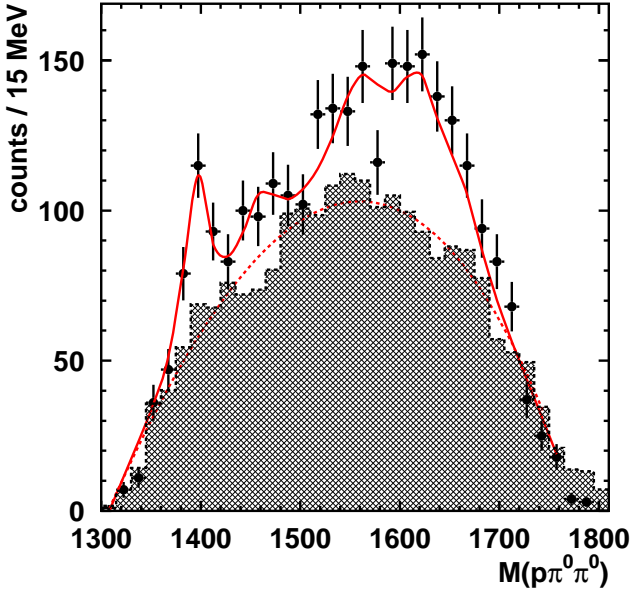}}  
\caption{(top) The invariant mass $M(p\pi^0\pi^0)$ versus the invariant mass
  $M(\pi^0\pi^0)$, requiring $M(p\pi^0)$ to be between 1160-1220 MeV. The
  incident photon energy is 2000-2300 MeV. The lines show the cut on
  $(\pi^0\pi^0)$ invariant mass for the y-projection. (bottom) The projection
  onto $M(p\pi^0\pi^0)$ axis. The fit function is composed of a polynomial
  background and Breit-Wigner resonance shapes including $\Sigma^*$(1385),
  $\Sigma^*$(1460), $\Sigma^*$(1560), $\Sigma^*$(1620), $\Sigma^*$(1660) and
  $\Sigma^*$(1670) resonances with parameters given by~\cite{pdg}. The shaded
  area represents the background in the sideband region of $K^0$, as marked by
  the dashed lines on the 2-dimensional plot above. \label{sstar}}
\end{figure}

\begin{figure}  
\resizebox{0.5\textwidth}{!}{%
  \includegraphics{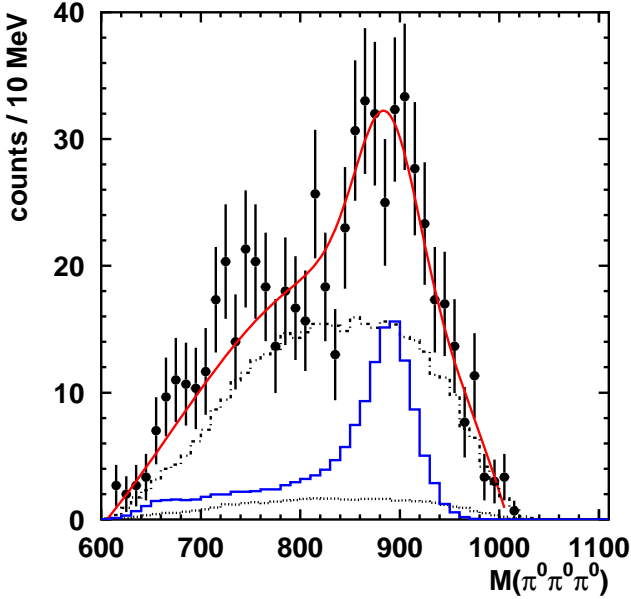}} 
\caption{(Color online) The experimental invariant mass (full black points)
  $M(\pi^0\pi^0\pi^0)$ for the incident photon 
  energy 2000-2150 MeV. The full fit is shown (red solid curve). The
  background is described by the contributions from higher $\Sigma^*$ states
  (dashed-dotted curve, and the 3-body phase space (dotted curve). The signal
  shape, including the combinatorial background, is given by the simulation
  (histogram). \label{ksbg}}
\end{figure}

From the $p4\pi^{0}$ final state, the $\Sigma^+$, $K^0$, and $K^{*0}$, are
reconstructed. The p$\pi^{0}$ and 3$\pi^{0}$ combinations are shown in
fig.~\ref{eta}b under the condition that the invariant mass of the 2$\pi^{0}$
from the 3$\pi^{0}$ on the x-axis should be: 470 MeV $< M(\pi^0 \pi^0) <$ 520
MeV. This cut has been applied to reconstruct the $\Sigma^+$ and $K^{*0}$
which require kaons to be selected (reactions (2), (3) and (4)). A peak at
$M(\pi^0\pi^0\pi^0)$ = 896 MeV and $M(p\pi^0)$ = 1189 MeV shows correlated
$K^*\Sigma$ production. Figures ~\ref{sigma}a  and ~\ref{sigma}b present the
projections on the $p\pi^{0}$ and $\pi^{0}\pi^{0}\pi^{0}$ invariant mass axes,
respectively. The spectrum in~\ref{sigma}a is fitted by a combination of
polynomial and Gaussian functions and shows a clear peak at 1189$\pm$2.0 MeV,
corresponding to $\Sigma^{+}$. The fit yields a resolution $\sigma$=15$\pm$3.1
MeV. The $\pi^0\pi^0\pi^0$ invariant mass spectrum in ~\ref{sigma}b, a
projection on the x-axis of the two-dimensional plot in fig.~\ref{eta}b with a
cut on $M(p\pi^0)$ in 1160-1220 MeV, shows a
peak around 896 MeV, corresponding to $K^{*0}$. Also after applying the cuts a
considerable background remains. The background in this spectrum is very
complex and is discussed later in this section. The $\pi^{0}\pi^{0}$ invariant
mass from six different combinations is shown in fig.~\ref{sigma}c. All
$\pi^0\pi^0$ combinations have been taken into account requiring that one of
the other two $\pi^0$'s and a proton have an invariant mass between 1160 and
1220 MeV, because the kaons are always produced with the $\Sigma^+$. The peak
at 496 MeV corresponds to $K^{0}$ mesons from one of the reactions (2), (3) or
(4). Higher lying hyperon  states $\Sigma^{*}$ could contribute to $K^{0}$
production via: $\gamma p \rightarrow K^{0}\Sigma^{*} \rightarrow
K^{0}\pi^{0}\Sigma^{+} \rightarrow (\pi^{0}\pi^{0})(\pi^{0}p\pi^{0})$, where
$\Sigma^{*}$ could be $\Sigma^{*}(1385)$ or higher lying $\Sigma^{*}$ states
decaying into $\pi^0 \Sigma^+$(1189). The threshold for the excitation of
$\Sigma^*$(1385) is $E_{\gamma}$=1400 MeV. The reactions on the proton
with neutral mesons only, such as $\gamma p \rightarrow
K^{0}\pi^{0}\Sigma^{+}$, excludes $\Lambda$'s as intermediate states. Requiring
the invariant mass $M(p\pi^0)$ to be close to the $\Sigma^+(1189)$ mass,
namely between 1160-1220 MeV, figure~\ref{sstar}(top) shows a plot of
$M(p\pi^0\pi^0)$ versus $M(\pi^0\pi^0)$ for incident photon energies of
2000-2300 MeV. The $\Sigma^{+}$ cut for this plot is important to
reconstruct the reaction (3), where $\Sigma^{*+}$ decays in
$\pi^{0}\Sigma^{+}$. The vertical band around 500 MeV on the x-axis is from the
$K^0$ events (cf. figure~\ref{sigma}c). The projection onto the y-axis, with a
cut on $M(\pi^0\pi^0)$ between 470 and 520 MeV is plotted in
fig.~\ref{sstar}(bottom), which corresponds to a $\pm
2.5\sigma$ cut on the $K^0$ invariant mass. This cut is shown with solid
vertical lines. The non $\Sigma^*$ background (shaded area in
fig.~\ref{sstar}) can be determined from side band cuts - shown with the
dashed vertical lines on figure~\ref{sstar}(top). Left and right from the
kaon peak, in the invariant mass range 445-470 MeV and 520-545 MeV, the
y-projection of the two-dimensional plot(fig.~\ref{sstar}) shows no $\Sigma^*$
peak (shaded area on fig.~\ref{sstar}(bottom).  The full
spectrum is fitted with a polynomial background and six Breit-Wigner resonance
shapes representing the $\Sigma^*$(1385), $\Sigma^*$(1460), $\Sigma^*$(1560),
$\Sigma^*$(1620), $\Sigma^*$(1660), $\Sigma^*$(1670) with parameters given by
~\cite{pdg}. The position and width of the resonances are taken from
~\cite{pdg} and the strengths of the corresponding Breit Wigner are taken as
free parameters. As it can be seen the polynomial background (dashed line on
fig.~\ref{sstar}(bottom)) is in good agreement with the background which we
got from the sidebands (shaded area). Apart from the $\Sigma^*$(1385) no
detailed information on the other $\Sigma^*$ resonances can be extracted due to
overlap. The estimated production cross section of the $\Sigma^{*}(1385)$
resonance is around 0.7$\pm$0.3 $\mu b$ at 1.85 GeV, which is of the same
order as observed for the $K\Sigma^{*}$ channel in ~\cite{wieland,guo}.\\
Monte Carlo studies have been performed to understand the $K^{*0}$ background
~\cite{me}. We assumed that the main contribution to the background in
$K^{*0}$ spectra is mainly caused by 4$\pi^{0}$ sequential resonance decays
and other channels leading to the same final state. The production of higher
lying hyperon resonances is hereby of particular importance. Since the final
particles produced by $K^0\Sigma^{*}$  and $K^{*0}\Sigma$ are the same, higher
lying hyperon states contribute to the $M(\pi^0\pi^0\pi^0)$ spectra as
well. To investigate this contribution we have simulated the reaction $\gamma
p \rightarrow K^{0}\Sigma^{*+}(1385)$ and the population of higher
$\Sigma^{*+}$ resonances which decay into $\pi^0\Sigma^+$(1189). The 3$\pi^0$
invariant mass calculated from the $K^0$ decay pions and the additional pion
from the $\Sigma^*$ decay contributes to the background in the $K^{*0}$
spectra as shown in figure~\ref{ksbg}. The contribution of $\Sigma^{*}$(1385)
and higher $\Sigma^{*}$ resonances to the background has been normalized to the
experimentally observed $\Sigma^{*}$ yields. The
background below the $K^{*0}$ signal is composed of the $\Sigma^{*}$ decay
contributions and a 3 body phase space part. The full fit of the experimental
$M(\pi^0\pi^0\pi^0)$ spectrum, shown in figure~\ref{ksbg}, has been done using
the $K^{*0}$ signal including the simulated combinatorial background, the
background from $\Sigma^*$, the 3-body phase space and adjusting their
relative magnitudes.\\ 
\begin{table}
\caption{$\Sigma^{+}$, $K^{0}$, $K^{*0}$ and $\Sigma^{+*}$ counts 
 in different incident photon energy bins (determined by integration of
  the respective angular distributions)}
\label{tab:1}
\begin{center}
\begin{tabular}{ccccc}
\hline\noalign{\smallskip}
$E_{\gamma}$ (MeV)&$\Sigma^{+}(1189)$&$K^{0}$(without $K^*$)&$K^{*0}$&$\Sigma^{*+}(1385)$\\      
\hline \hline
1400-1500& 16$\pm$4& 18$\pm$5& - & -\\
\noalign{\smallskip}\hline
1500-1700&77$\pm$11 &70$\pm$13&-& -\\
\noalign{\smallskip}\hline
1700-1850&92$\pm$16&95$\pm$19& - & -\\
\noalign{\smallskip}\hline
1850-2000&244$\pm$34&152$\pm$39&92$\pm$20&30$\pm$10\\
\noalign{\smallskip}\hline
2000-2150&340$\pm$35&240$\pm$40&100$\pm$20&111$\pm$25\\
\noalign{\smallskip}\hline
2150-2300&240$\pm$34&146$\pm$37&94$\pm$15&30$\pm$10\\
\noalign{\smallskip}\hline 
2300-2500&290$\pm$38&181$\pm$40&109$\pm$16&16$\pm$10\\
\hline \hline
sum& 1299$\pm$73&902$\pm$81&395$\pm$36&187$\pm$30\\                                   
\hline \hline
\end{tabular}
\end{center}
\end{table}
The numbers of the identified $\Sigma^{+}$, $K^{0}$, $K^{*0}$ and
$\Sigma^{+*}$ are listed in Table~\ref{tab:1} for different photon energy bins.
The statistical error of the cross section data has been
estimated by $\Delta S =\sqrt{S+2B}$, where S are the counts in
the signal and B are the counts in the background underneath the signal. 
\begin{figure} 
\resizebox{0.5\textwidth}{!}{%
  \includegraphics{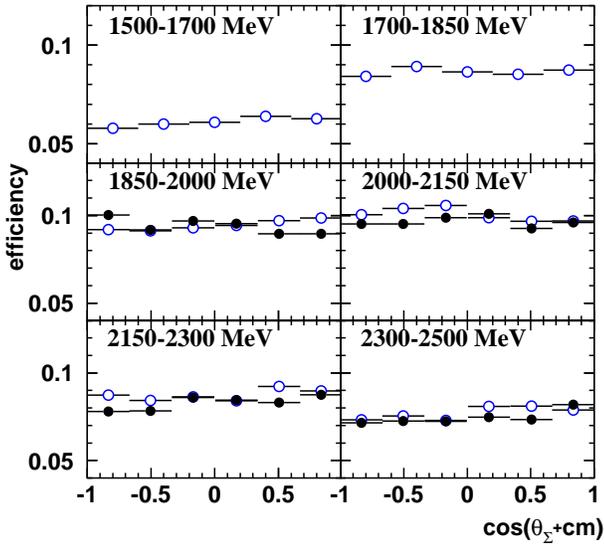}}
\caption{Total detection efficiency for the reactions $\gamma p \rightarrow
  K^{0}\pi^0\Sigma^{+}$ (open circles) and $\gamma p \rightarrow
  K^{*0}\Sigma^{+}$ (full circles) as a function of the center-of-mass angle
  of the $\Sigma^{+}$ for different incident photon energy bins. \label{eff1}}
\end{figure}
\section{Results and discussion}
\label{sec:6}
\subsection{Absolute reaction cross sections}
\label{sec:6.1}
This section describes how the absolute reaction cross sections are
determined. The essential ingredients are the reaction yields, the detection
efficiency of the individual final states and the photon flux.
Due to the almost 4$\pi$ coverage, the detection efficiency for the hyperon
final states is practically independent of the production angle. This is
illustrated for both reactions $\gamma p \rightarrow
(K^{0}\pi^{0}\Sigma^{+})_{n.r.}$ and $\gamma p \rightarrow K^{*0}\Sigma^{+}$
in fig.~\ref{eff1}, where the total efficiency is shown, including
geometrical acceptance and the detector efficiency. The total efficiency was
determined with a GEANT-based Monte Carlo simulation. The simulated events
were evenly distributed over the available phase space and analyzed using the
same  event selection criteria, kinematic fit, applied cuts, thresholds and
trigger conditions as for the experimental data. The resulting efficiency
varies slightly between 7-10\% with the incoming photon energy; its angular
dependence is very small.\\
Uncertainties in the reconstruction of hyperons and vector mesons have been
studied. By varying the fit conditions in order to achieve a consistent
description of the background in different kinematical regions, an error of 3\%
-15\% is deduced.\\
The photon flux through the target is determined by counting the
photons reaching the $\gamma$ intensity detector in coincidence with electrons
registered in the tagger system. This provided an absolute normalization for
all measurements. The main point is the  
accurate determination of the efficiency of the tagging system, as defined by
the probability to identify the corresponding photon in the photon beam for
each detected electron in the tagger. The $\gamma$ detector has
almost 100\% photon detection efficiency. By
comparing the number of electrons in the tagging system in
coincidence with the number of counts in the $\gamma$ detector, the
tagging efficiency has been determined to vary between 64 to 74 \%
for different beamtimes. The systematic uncertainty in
the cross sections, caused by the photon flux determination has been checked
by measurements of known reactions, such as $\gamma p\rightarrow p \eta$, and
is estimated to be 5\%-15\% depending on the photon energy.  
\subsection{Differential cross section}
The differential cross sections are calculated from the number of events
identified in the respective channel using:
\begin{equation}
\frac{d\sigma}{dcos(\theta^{cm})}{=}\frac{N_{\Sigma^{+}}}{A_{\Sigma^+
    \rightarrow p\pi^0}}\frac{1}{N_\gamma \rho_{t}} \frac{1}{\Delta
    cos(\theta)}\frac{\Gamma_{total}}{\Gamma_{\Sigma^+ \rightarrow p\pi^0}}
\label{cross}
\end{equation}
$N_{\Sigma^+}$ ($N_{K^*}$) are the counts of $\Sigma^+$ ($K^*$) determined in
different angle and energy bins as described in Sect.~\ref{sec:5}.\\ 
$A_{\Sigma^+ \rightarrow p\pi^0}$ ($A_{K^* \rightarrow K^0\pi^0}$) is the
efficiency determined as described in Sect.~\ref{sec:6.1};\\
$N_\gamma$ is the number of primary photons in the respective energy bin
determined as described in Sect.~\ref{sec:6.1};\\
$\rho_t$ is the target area density for the
LH$_2$ target used in this experiment;\\
$\Delta cos(\theta^{cm})$ is the angle bin width of the angular
distributions.\\ 
$\frac{\Gamma_{\Sigma^+ \rightarrow p\pi^0}}{\Gamma_{total}}$
($\frac{\Gamma_{K^* \rightarrow K^0\pi^0}}{\Gamma_{total}}$) is the branching
ratio of the reaction $\gamma p \rightarrow K^{0}\pi^{0}\Sigma^{+}$  ($\gamma
p\rightarrow K^{*0}\Sigma^{+}$) respectively; $\pi^0$'s were identified
via their decay into 2$\gamma$ which has a relative branching ratio of 98.798\%.\\
\begin{figure} [h]
\resizebox{0.5\textwidth}{!}{%
  \includegraphics{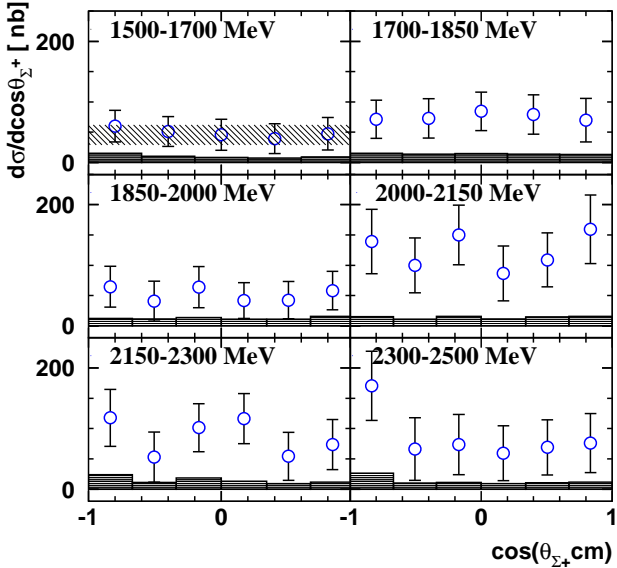}}
\caption{Differential cross section $d\sigma$/d$cos(\theta_{\Sigma^{+}}^{cm})$
  for $\gamma p\rightarrow K^{0}\pi^{0}\Sigma^{+}$. Hereby, the contribution
  from $K^{*0}\Sigma$ final state has been removed. In the range of energies
  below the threshold of $K^{*0}$ production the theoretical band for the
  angular distribution of $\gamma p\rightarrow K^{0}\pi^{0}\Sigma^{+}$
  ~\cite{doering} is the shaded area on the first picture. The systematic
  errors are shown in the boxes on the bottom of each picture.
\label{wqdiff1}}
\end{figure}
\begin{figure} 
\resizebox{0.5\textwidth}{!}{%
  \includegraphics{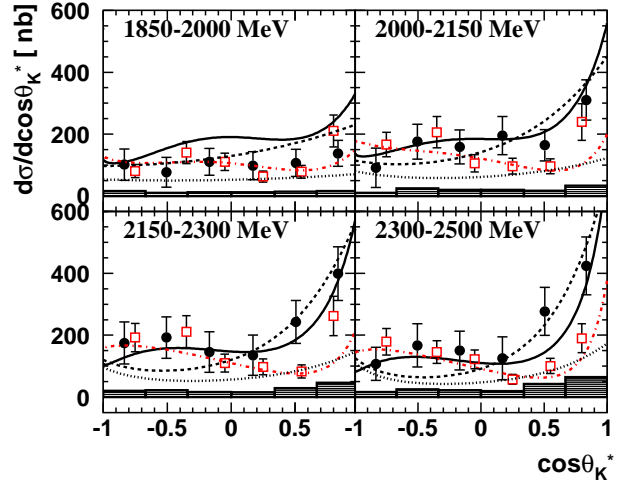}}
\caption{(Color online) Differential cross section $\gamma p\rightarrow
  K^{*0}\Sigma^{+}$ (full circles). The empty squares are the CLAS
  experimental data~\cite{Hleiq}. The solid curve represents the theoretical
  calculations for the  $\gamma p \rightarrow K^{*0}\Sigma^{+}$ reaction
  in~\cite{zhao,zhao_up} with parameters $a = 2,7$ and $b = -1,7$. The dashed
  and dotted curves denote the calculations in ~\cite{Oh} for model II and I
  respectively. The dashed-dotted curve represents calculations
  in~\cite{zhao_up} with free parameters $a = -2,2$ and $b = 0,8$. The
  grey band on the botom represents the systematic uncertainty. \label{wqdiff}}
\end{figure}
The angular distributions for the reactions $\gamma p\rightarrow
K^{0}\pi^{0}\Sigma^{+}$ and $\gamma p\rightarrow
K^{*0}\Sigma^{+}$ are shown in figures ~\ref{wqdiff1} and ~\ref{wqdiff},
respectively. The results are plotted as a function of
$cos(\theta_{\Sigma^{+}}^{cm})$ and as a function of
$cos(\theta_{K^{*}}^{cm})$  respectively. For incident photon energies higher
than 1850 MeV, above the threshold for the $K^{*0}$ production, the cross
section for the $\gamma p\rightarrow K^{0}\pi^{0}\Sigma^{+}$ has been
extracted from the difference between $\Sigma^{+}$ and $K^{*0}$ yields, i.e.
$\sigma_{\Sigma K\pi}= \sigma_{\Sigma} - \sigma_{\Sigma K^*}$ .\\ 
In the energy range 1500-1850 MeV, below the $K^{*}$ production threshold, the
differential cross section is almost flat. These measurements are in good
agreement with the theoretical prediction of ~\cite{doering}, based on the
model of ~\cite{oset1}, as shown in the first picture in
figure~\ref{wqdiff1}. The flat angular distribution indicates dominant
s-channel production which is a genuine prediction of a chiral dynamical
calculation based on the dominance of the $\Delta(1700)$ in the entrance
channel, plus the coupling of this resonance to $K \Sigma^*(1385)$. For the
energy region 1500-1850 MeV, where the $K^{*}$ production is
energetically not possible, the reconstructed $\Sigma^{+}\pi^{0}K^{0}$ events
could come from a $\Delta^{*}$ resonance which decays subsequently in
$\Sigma^{*}(1385) K$ and $\Sigma^{+}\pi^{0}K^{0}$. For energies higher than
1850 MeV, an additional contribution from the reaction $\gamma p\rightarrow
K^{*0}\Sigma^{+}$ is possible. These two contributions can be separated
experimentally through the $K^{0}\pi^{0}$ mass spectrum which exhibits a sharp
peak at the position of the $K^{*0}$ meson (fig.~\ref{ksbg}). The observed
counts in the peak at different bins of $cos(\theta_{K*}^{cm})$ are used for
the differential cross section determination (fig.~\ref{wqdiff} full
circles).\\ 
The differential cross  section d$\sigma$/d$cos(\theta_{K*}^{cm})$
for the reaction $\gamma \rightarrow K^{*0}\Sigma^{+}$  shows a rise in the
forward direction when plotted vs the $K^{*0}$ production angle
(fig.~\ref{wqdiff}). Production of the $K^{*0}$ meson via $t$-channel exchange
seems to play an important role in the reaction dynamics. These results are
compared with the updated calculations from the model in ~\cite{zhao_up},
using the free parameters $a = 2.7$ and $b = -1.7$, which describe the
universal couplings for the vector and tensor part in the quark vector-meson
interaction. These parameters are different from those given in ~\cite{zhao},
which were derived from $\omega$ meson photoproduction based on SU(3)
symmetry. Our data are compared
to the experimental data from the CLAS collaboration~\cite{Hleiq} which also
show an enhancement in the forward direction. At
backward angles we do not observe any rise in the cross section. There is a 
discrepancy to the CLAS data for the forward angle bins and at incident
photon energies higher than 2150 MeV. Our
experimental data are consistently higher at angles with
$cos(\theta_{K*}^{cm}) > 0.5$ in the lower two pictures in
fig.~\ref{wqdiff}.\\  
A comparison to another theoretical
model assuming $\kappa$(800) meson exchange ~\cite{Oh} is also shown. The
theoretical curves of the so-called model II provide a reasonable
agreement with the experimental data which is not the case for model I. The
main difference between models I and II of
ref.~\cite{Oh} is in the form and the strength of the form factor of the
$\kappa$(800) meson.\\
\begin{figure} [h]
\resizebox{0.5\textwidth}{!}{%
  \includegraphics{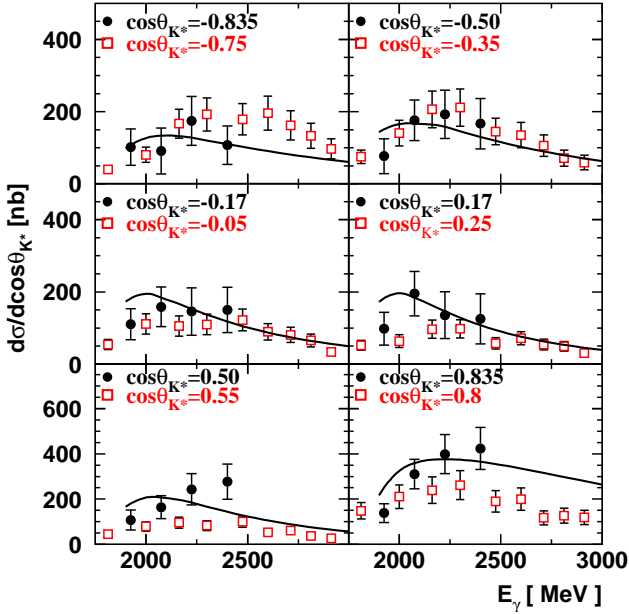}}
\caption{Differential cross sections for the reaction $\gamma p \rightarrow
  K^{*0}\Sigma^{+}$ as a function of the incident photon energy for the
  angular bins shown in fig.~\ref{wqdiff}. Our experimental data (full
  circles) are compared to the CLAS data (empty squares)~\cite{Hleiq}. The
  solid curve refers to theoretical prediction for the $\gamma
  p\rightarrow K^{*0}\Sigma^{+}$ reaction in~\cite{zhao_up}.\label{wqene}}
\end{figure}
\begin{figure} 
\resizebox{0.5\textwidth}{!}{%
    \includegraphics{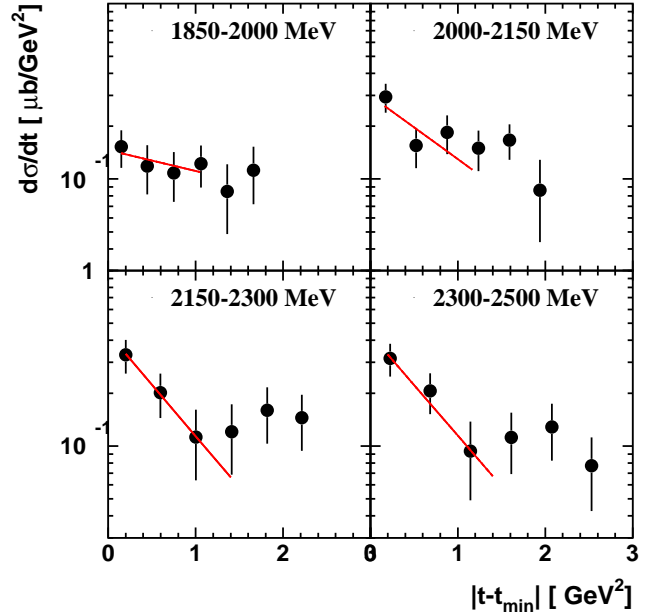}}
\caption{Differential cross section $d\sigma$/dt for $\gamma p\rightarrow
  K^{*0}\Sigma^{+}$ . For small values of momentum transfer the
  data are fitted by an exponential function $e^{a+b|t-t_{min}|}$.
  \label{tchan}}
\end{figure} 
The energy dependence of the differential cross sections for the reaction
$\gamma \rightarrow K^{*0}\Sigma^{+}$ is presented in fig.~\ref{wqene} for six
angular bins and compared with the experimental data from CLAS. The
peaking in the forward direction is more pronounced than in the CLAS data as
it can be seen in the lowest two pictures in figure~\ref{wqene}.
\subsection{t-Scaling}
\begin{figure}  
\resizebox{0.5\textwidth}{!}{%
    \includegraphics{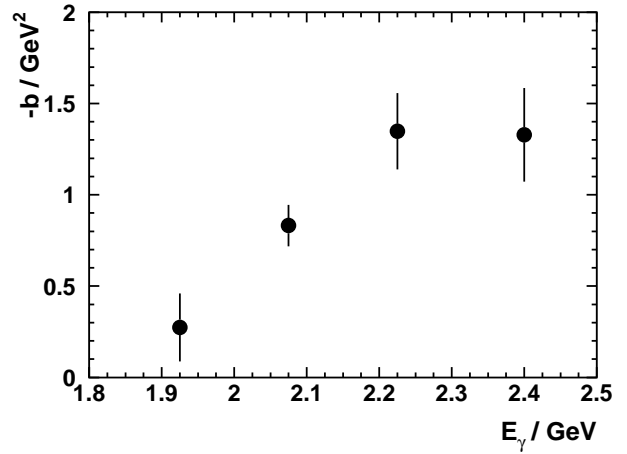}}
\caption{The slope parameter $-b$ of $e^{a+b|t-t_{min}|}$ as a function of
  the beam energy $E_{\gamma}$. The error bars include the
  uncertainties on the slope fits shown in fig.~\ref{tchan}. \label{slope}}
\end{figure}

The forward peaking of the $K^{*0}\Sigma^{+}$ cross section suggests that
there is a contribution to the reaction mechanism from $t$-channel exchange. To
test this idea the differential cross section
d$\sigma$/d$cos(\theta_{K^*}^{cm})$ as function of $cos(\theta_{K^*}^{cm})$ was
converted to d$\sigma$/$dt$ as a function of $t-t_{min}$.
The variable $t$ is the Mandelstam invariant that gives the 4-momentum
squared of the exchange particles. Since the momentum transfer is
limited by kinematics, $t$ lies between $t_{min}$ and $t_{max}$, given by: \\
\begin{equation}
t_{min,max}{=}\biggl[\frac{m_{K*}^{2}}{2\sqrt{s}}\biggr]^{2}-\biggl[\frac{s-m_{p}^{2}}{2\sqrt{s}}\mp \sqrt{\frac{(s+m_{K*}^{2}-m_{p}^{2})^{2}}{4s}-m_{K*}^{2}}\biggr]^{2}
\label{fitt1}
\end{equation}
The cross  section d$\sigma$/$dt$ is shown in
fig.~\ref{tchan}. The straight lines represent fits of
$e^{a+b|t-t_{min}|}$ in the region below 1.0 $GeV^2$ of $|t-t_{min}|$. The
slope parameter $b$ has a negative value. We
plot the slope parameter $-b$ as a function of the incident photon energy
$E_{\gamma}$ in fig.~\ref{slope}. It can be seen that the slope parameter
rises with the photon energy. This is an indication for an increasing
contribution to $K^{*0}$ production via $t$-channel exchange as
predicted in~\cite{zhao}.
\subsection{Total cross section}
\begin{figure} 
\resizebox{0.55\textwidth}{!}{%
      \includegraphics{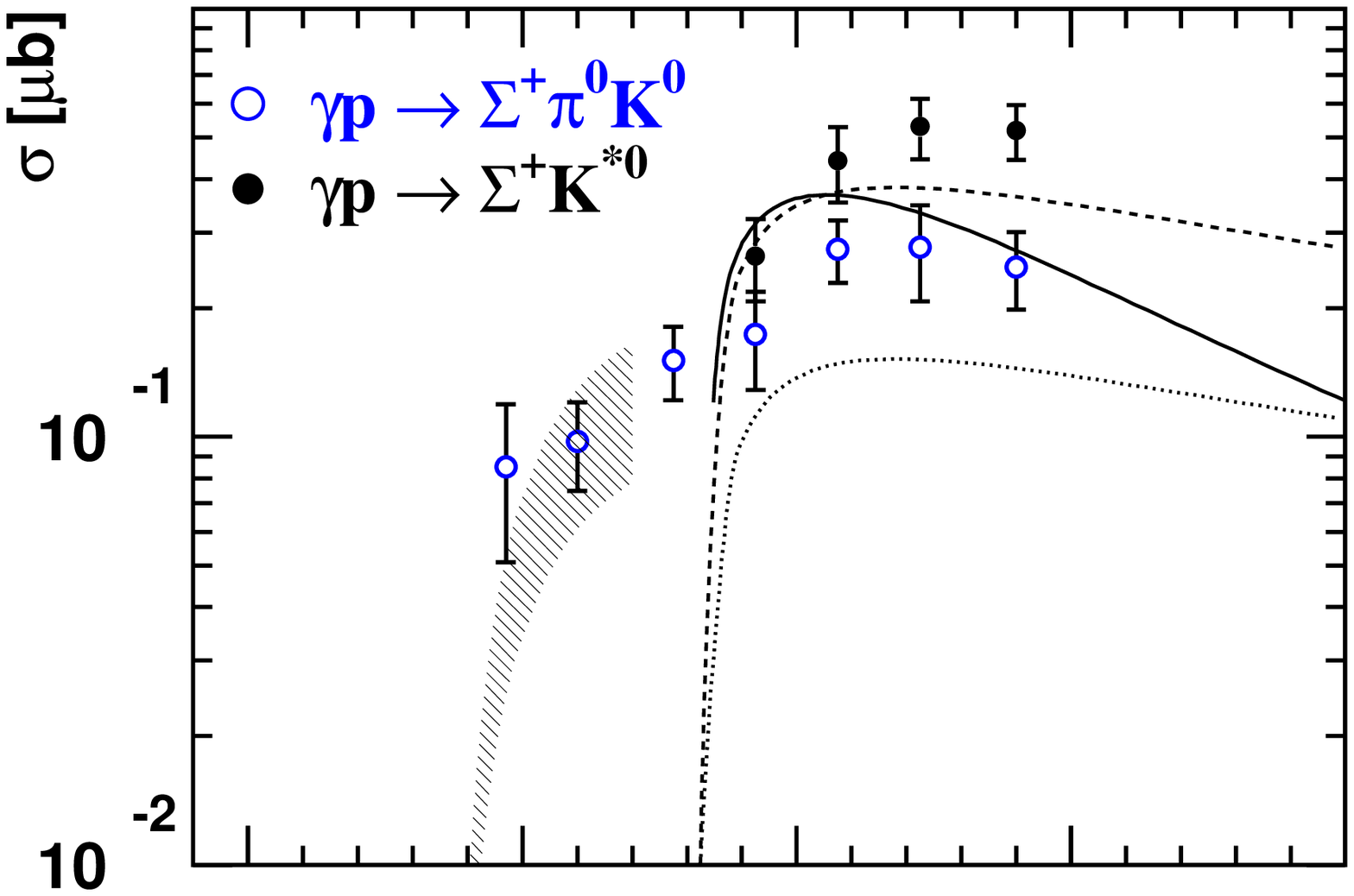}}
\resizebox{0.55\textwidth}{!}{%
      \includegraphics{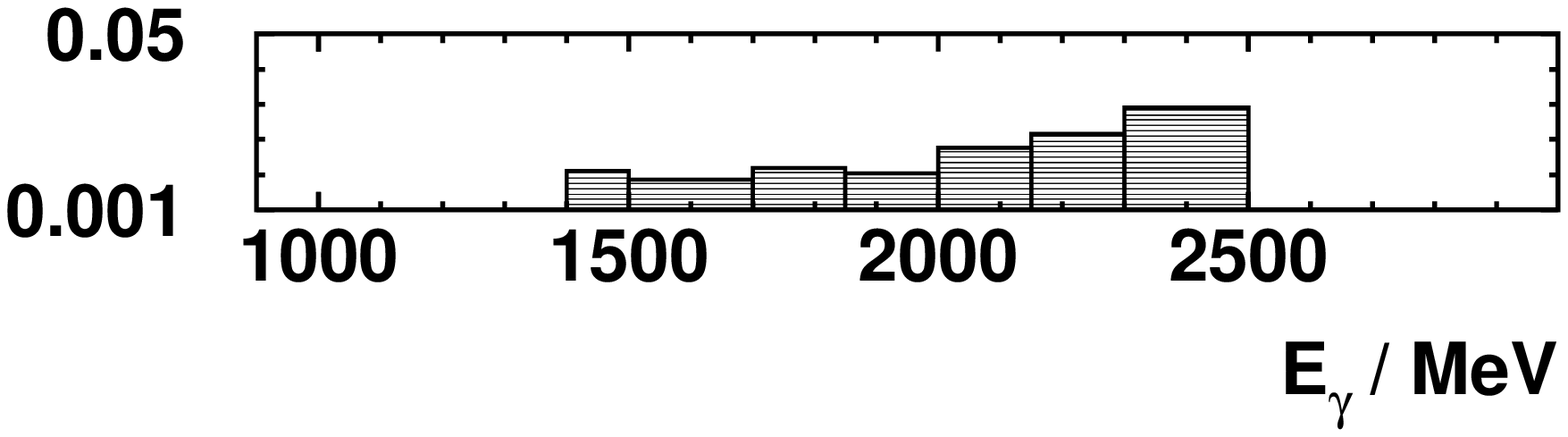}}
\caption{(Color online) The energy dependence of the total cross sections for
  $\gamma p\rightarrow K^{0}\pi^{0}\Sigma^{+}$ (empty circles) and $\gamma
  p\rightarrow K^{*0}\Sigma^{+}$ (filled circles). The shaded area represents
  a band for the predicted values from the theoretical calculations of the
  $\gamma p\rightarrow K^{0}\pi^{0}\Sigma^{+}$  reaction in ~\cite{oset1}.
  The solid curve refers to theoretical prediction for the $\gamma
  p\rightarrow K^{*0}\Sigma^{+}$ reaction in~\cite{zhao}. The dashed and dotted
  curves are the calculations with $\kappa$ meson exchange~\cite{Oh} in the
  $t$-channel, model II and model I, respectively. The systematic errors are
  shown as an error band on the bottom. \label{total}}
\end{figure}

The total cross sections for the $\gamma p\rightarrow K^{0}\pi^{0}\Sigma^{+}$
and $\gamma p\rightarrow K^{*0}\Sigma^{+}$ reactions are shown in
fig.~\ref{total}. 
The cross sections agree within errors with those determined by integrating
over the differential cross sections.
The experimental cross section for the $\gamma p\rightarrow K^{*0}\Sigma^{+}$
channel are in reasonable agreement with theoretical predictions of
reference~\cite{zhao} with the assumption of a $t$-channel $K^{0}$
exchange.\\  
The predicted total cross section for $K^{0}\pi^{0}\Sigma^{+}$ in
~\cite{oset1} is given by a band (the shaded area in
fig.~\ref{total}). The good agreement between theory and experiment suggests
that the dynamics used in the model is reasonable for the corresponding energy
region.
The cross section for $\gamma p\rightarrow K^{0}\pi^{0}\Sigma^{+}$ rises with
increasing photon energy. Above 1850 MeV the dominating channel for
$\Sigma^{+}$ production is obviously $\gamma p\rightarrow
K^{*0}\Sigma^{+}$. It is about factor 2 larger than the cross section of the
$\gamma p\rightarrow K^{0}\pi^0\Sigma^{+}$ reaction which includes excited
hyperon decays into the $\Sigma^+(1189)$. 
\section{Summary}
We have reported measurements of differential and total cross sections for the
$\gamma p\rightarrow K^{0}\pi^{0}\Sigma^{+}$ and $\gamma p\rightarrow
K^{*0}\Sigma^{+}$ reactions. The experimental data have been compared with the
available theoretical predictions. At low incident photon energies below the
$K^{*0}$ production threshold ($E_{\gamma}$ $<$ 1850 MeV) only the
$K^{0}\pi^{0}\Sigma^{+}$ channel is energetically possible, exhibiting a flat
angular distribution, dominated by s-channel production which is the
prediction based on the dominance of the $\Delta(1700)$, subsequently decaying
into $K\Sigma^{*}(1385)$.\\ 
For energies above 1850 MeV the $\Sigma^{+}$ production is mainly associated
with $K^{*0}$ production. The angular distributions for the reaction
$\gamma p\rightarrow K^{*0}\Sigma^{+}$ show a rise in the forward direction of
the vector meson which indicates a $t$-channel exchange contribution to
the reaction mechanism. It is, however, not yet possible to make an explicit
conclusion about the exchanged particle, due to the significant contribution
to the $K^{0}\pi^{0}\Sigma^{+}$ channel from the $\Sigma^{*+}$ production.\\ 
The presented data provide valuable confirmation of the theoretical
predictions concerning the reaction mechanism of the $\gamma
p\rightarrow \pi^{0}K^{0}\Sigma^{+}$ and $\gamma p\rightarrow
K^{*0}\Sigma^{+}$ reactions. $\Sigma^{*}(1385)$ and higher $\Sigma^{*}$
resonances have been observed in the $\Sigma^+\pi^0$ decay channel. On the
basis of existing data we have estimated the production cross section of the
$\Sigma^{*}(1385)$ resonance. A corresponding value for the cross section of
the higher $\Sigma^{*}$ states photoproduction can not be given because of
the overlap of these resonances. Further
experiments are required to study the higher $\Sigma^*$ resonances.
Polarisation experiments will be needed to clarify
details of the strangeness production process.
\label{sec:8}
\section{Acknowledgements}
We thank the accelerator group of ELSA as well as the technicians of the
participating institutions for their outstanding support. We acknowledge
illuminating discussion with Kenneth Hicks on details of the data analysis. 
Useful discussions with E. Oset and M. D\"oring and also Q. Zhao on the
theoretical interpretation of the data are also highly acknowledged.
This work was supported by Deutsche Forschungsgemeinschaft SFB/TR16 and
Schweizerischer Nationalfond.

\end{document}